\documentclass{article}
\usepackage{
dcase2020,
amsmath,
amsfonts,
graphicx,
url,
times,
booktabs,
tabularx,
xurl,
cite,
array,
comment
}

\newcommand{\bm}[1]{\boldsymbol{#1}}

\newlength{\tablelength}
\newcolumntype{C}[1]{>{\centering\arraybackslash}p{#1\tablelength}}
\newcolumntype{L}[1]{>{\arraybackslash}p{#1\tablelength}}

\newcommand{\mysection}[1]{
\vspace{-3pt}
\section{#1}
\vspace{-3pt}
}

\newcommand{\mysubsection}[1]{
\vspace{-3pt}
\subsection{#1}
\vspace{-2pt}
}

\title{Effects of Word-frequency based Pre- and Post- Processings\\
for Audio Captioning}

\name{Daiki Takeuchi, Yuma Koizumi, Yasunori Ohishi, Noboru Harada, and Kunio Kashino}
\address{
NTT Corporation, Japan
}

\begin{document}

\ninept
\maketitle

\begin{sloppy}

\begin{abstract}

The system we used for Task 6 (Automated Audio Captioning) of the Detection and Classification of Acoustic Scenes and Events (DCASE) 2020 Challenge combines three elements, namely, data augmentation, multi-task learning, and post-processing, for audio captioning.
The system received the highest evaluation scores, but which of the individual elements most fully contributed to its performance has not yet been clarified. 
Here, to asses their contributions, we first conducted an element-wise ablation study on our system to estimate to what extent each element is effective. We then conducted a detailed module-wise ablation study to further clarify the key processing modules for improving accuracy.
The results show that data augmentation and post-processing significantly improve the score in our system.
In particular, mix-up data augmentation and beam search in post-processing improve SPIDEr by 0.8 and 1.6 points, respectively. 
\end{abstract}


\mysection{Introduction}
\label{sec:intro}


Automated audio captioning (AAC) is a crossmodal translation task in which input audio is translated into a description of the audio using natural language \cite{drossos2017automated, drossos2019clotho, wu2019audio, kim2019audiocaps, ikawa2019neural, koizumi2020transformer}.
Whereas automatic speech recognition (ASR) converts speech to text, AAC converts environmental sounds to text.
Generating meaningful captions for environmental sounds requires the incorporation of higher contextual information, including concepts, physical properties, and high-level knowledge, though such information is not necessarily needed for tagging scenes~\cite{mesaros2010acoustic, imoto2020sound}, events~\cite{barchiesi2015acoustic}, and conditions~\cite{koizumi2018unsupervised}.

For the AAC task of the Detection and Classification of Acoustic Scenes and Events (DCASE) Challenge 2020, Drossos et al. set up Task 6~\cite{dcase2020task6},
which prohibited the use of external resources and pre-trained models. 
Thus, participants were required to solve task-specific problems for audio captioning, rather than provide computational generic solutions using large amounts of training data or large-scale pre-trained models such as VGGish~\cite{vggish} and BERT~\cite{devlin2018bert}.
This is because, even though image or video captioning is being actively studied, AAC is still an emerging research field, and because there has not been an exhaustive study on methods for improving accuracy.

Here, we address three research questions in AAC. The first is how to augment the limited data for effective training.
Therefore, we discuss data augmentation that makes effective use of small amounts of data.
The second is how to decide the best description for given audio.
As described in \cite{ikawa2019neural,koizumi2020transformer}, information contained in audio signals can be much more ambiguous than that in images or videos. Moreover, the validity of the description generally depends on the situation or context as well as the sound itself. Therefore, we introduce a deep neural network~(DNN) architecture for estimating keywords and sentence length along with caption generation as multi-task learning.
The third is whether post-processing such as beam search decoding is effective in AAC. This is because it has been found that post-processing can correct obviously incorrect output in machine translation and image captioning.

Figure \ref{fig:system} shows an overview of our system which incorporates data augmentation, a DNN architecture with multi-task learning, and post-processing, as summarized in Table \ref {tab:method}. Although the system yielded the best score at the challenge, the effectiveness of each module has not yet been clarified.

This paper is organized as follows. In Section 2, we describe in detail the research questions that were raised through this challenge and explain our motivation for adopting each module. As described in Section 3, we carried out an ablation study. First, we roughly divide our system into three processing elements for ablation, (i) data augmentation, (ii) multi-task learning, and (iii) post-processing. Based on the result showing that (i) and (iii) mainly contribute to the accuracy, we then conducted a finer ablation study for each component module of (i) and (iii).
Our findings are summarized as follows:

\begin{table}[ttt]
\caption{The three elements and their component modules used in our system\cite{koizumi2020t6ntt}}
\label{tab:method}
\centering
\begin{tabular}{ @{\hspace{2pt}}L{0.35}@{} @{\hspace{5pt}}L{0.6}@{\hspace{2pt}} }
\toprule
(i) Data augmentation    & (a) Mix-up~\cite{zhang2018mixup} \\
                         & (b) TF-IDF-based word placement~\cite{xie2019unsupervised}\\
                         & (c) IDF-based sample selection \\
\midrule
(ii) Multi-task learning & (a) Sentence length estimation\ \\
                         & (b) Keyword estimation~\cite{koizumi2020transformer} \\
\midrule
(iii) Post-processing    & (a) Beam search decoding~\cite{koehn2009statistical, koehn2004pharaoh} \\
                         & (b) Test time augmentation~\cite{ayhan2018testtime}\\
\bottomrule
\end{tabular}
\vspace{-2pt}
\end{table}

\begin{figure}[t]
  \centering
\includegraphics[width=0.9\columnwidth]{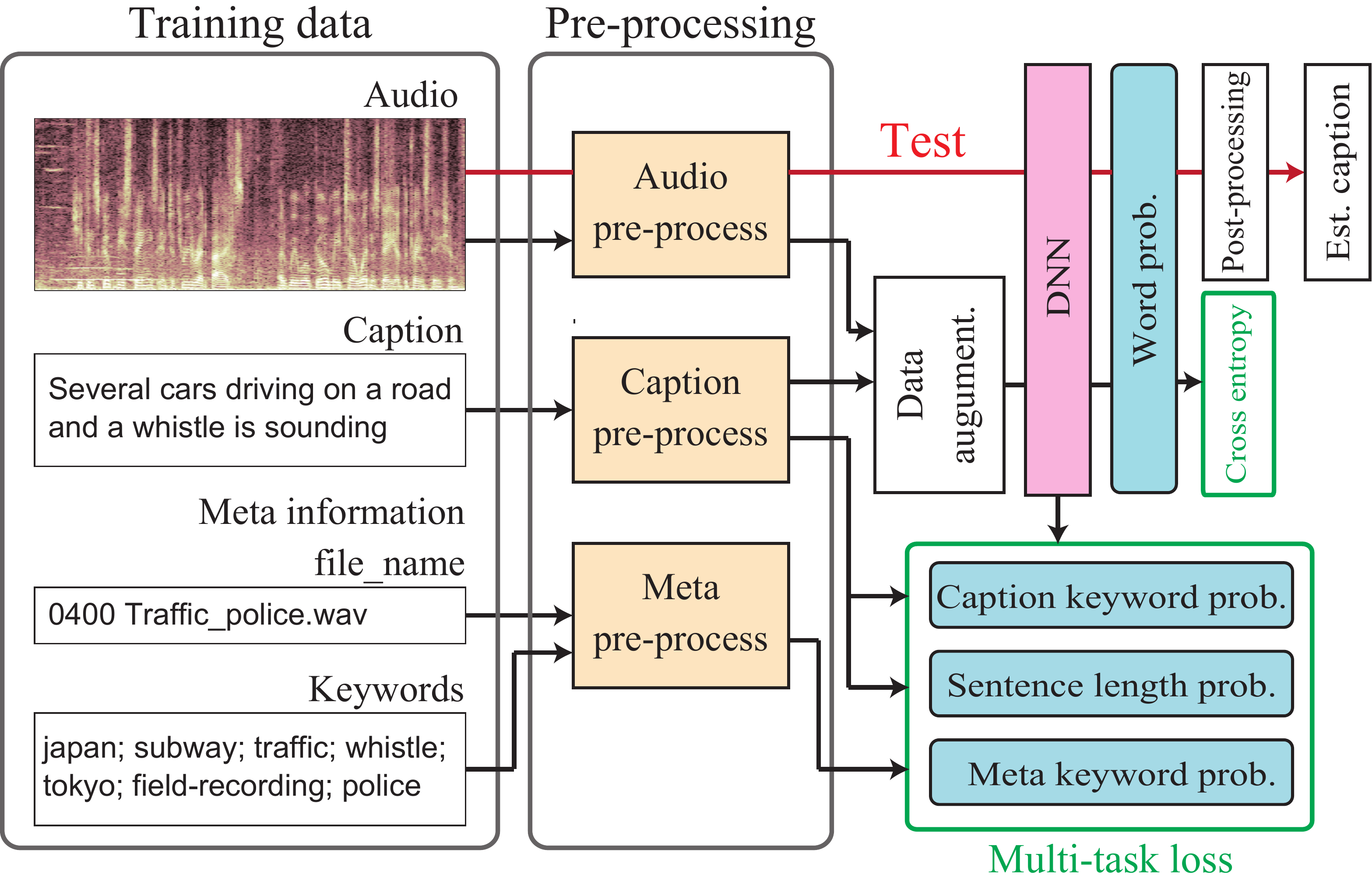} 
  \vspace{-10pt}
  \caption{System overview.}
  \label{fig:system}
  \vspace{-10pt}
\end{figure}

\begin{enumerate}
 \setlength{\parskip}{0cm} 
 \setlength{\itemsep}{0.1cm} 
\item
Data augmentation and post-processing significantly improved the performance of AAC.
\item
Multi-task learning did not improve the performance under the task rules, even though it was effective in more computationally rich scenarios, i.e., with pre-trained models \cite{koizumi2020transformer}.
\item
Mix-up data augmentation 
was effective as with acoustic event detection and acoustic scene classification.
\item
Beam search decoding was effective for AAC as well as other text generation tasks including ASR~\cite{seki2019vectorized} and image captioning~\cite{anderson2017guided}.
\end{enumerate}
\vspace{-2pt}


\mysection{System description}
\label{proposed}
This section describes our research questions and their solutions.
See our technical report for the implementation details.

\mysubsection{Research questions in AAC}

We identified three research questions in AAC:

\vspace{3pt}
\noindent
{\bf Data augmentation:}
In AAC, unlike other language processing tasks such as machine translation, it is difficult to collect training data from the Web. Therefore, the amount of training data available is inevitably less than that for other natural language processing tasks. In fact, there are only 14,465 training captions in the Clotho dataset \cite{drossos2019clotho}, whereas there are 36M in the WMT 2014 English-French dataset for machine translation.
This leads to the first research question: {\it how to augment the limited data for effective training}.

\vspace{3pt}
\noindent
{\bf Multi-task training for resolving indeterminacy:}
As described in our previous studies, AAC has two indeterminacy problems.
One is the indeterminacy in word selection \cite{koizumi2020transformer}, and the other is that in the degree of detail and sentence length \cite{ikawa2019neural}. The first problem arises because one acoustic event/scene can be described in several different ways using different sets of words, such as \{{\it car, automobile, vehicle, wheels}\} and \{{\it road, roadway, intersection, street}\}\cite{koizumi2020transformer}. The second problem appears because a sound can be explained briefly or in detail with a long expression, such as ``car sounds,'' or ``small cars and large trucks are driving on a roadway and they are making very loud engine noises'' \cite{ikawa2019neural}. Such indeterminacy leads to a combinatorial explosion of possible answers, making it almost impossible to estimate the best description and appropriately train AAC systems. 

\vspace{3pt}
\noindent
{\bf Post-processing:}
Post-processing such as beam search decoding is effective for improving accuracy. 
Post-processing in natural language processing can correct obviously incorrect output based on some explicit criteria.
In DNN-based text generation, some post-processing methods also modify the output of the DNN to improve accuracy.
Since AAC is a crossmodal text generation task with audio input, it is worth investigating whether post-processing is also effective in AAC and what kind of post-processing is effective.



%

\mysubsection{Data augmentation}

\vspace{3pt}
\noindent
{\bf IDF-based sample selection:} 
Since the objective metrics for captioning such as CIDEr \cite{vedantam2015cider} and SPIDEr \cite{liu2017improved} focus on the accuracy of infrequent words and idioms, we adopt two tricks for selecting the training samples based on IDF.

The first one is used for selecting an audio sample $\bm{x}$ from the training dataset.
First, we concatenate five ground-truth captions corresponding to each audio clip $\bm{x}$ in the training dataset and use the concatenated captions as a ``sentence''.
Then, we calculate the IDFs for all words in all sentences and the average IDF of each sentence.
Finally, each average IDF is normalized by the sum of the average IDFs. 
We regard the normalized IDF as the parameter of the categorical distribution and select $\bm{x}$ based on this probability.
Since data with a high average IDF contain low frequency words in the dataset, it is assumed that the data contain low frequency topics.
Increasing the probability of selecting low frequency topics is expected to prevent over-fitting on high frequency topics.


The second trick is used for selecting a ground-truth caption $\bm{w}$ from five captions corresponding to the selected $\bm{x}$.
The basic strategy is the same as in the first trick.
First, we calculate the IDFs of all words in the five captions. 
The document used here comprises the five captions, while the document used in the first trick consists of all sentences of the dataset.
Then, we calculate the normalized IDF and use that as the parameter of categorical distribution. The target caption $\bm{w}$ is selected based on this probability.
By increasing the probability that low frequency words are selected, we expect to avoid over-fitting on high frequency words.

\vspace{3pt}
\noindent
{\bf TF-IDF-based word placement:} 
Considering the limited number of sentences that can be used for training AAC systems, sentences in the dataset should be augmented in some way.
One simple data augmentation method is to replace the words randomly.
However, the meaning of sentences can change dramatically if informative words are replaced, and the correspondence between the caption and the audio clip will collapse. To cope with this problem, we need to control the words to be replaced. Thus, we adopt the TF-IDF based word placement method ~\cite{xie2019unsupervised} to augment texts.
This method avoids replacing informative words by limiting the words to be replaced to those with low TF-IDF and augments text data without largely changing the meaning of sentences.



\begin{figure}[t]
 \centering
\includegraphics[width=0.9\columnwidth]{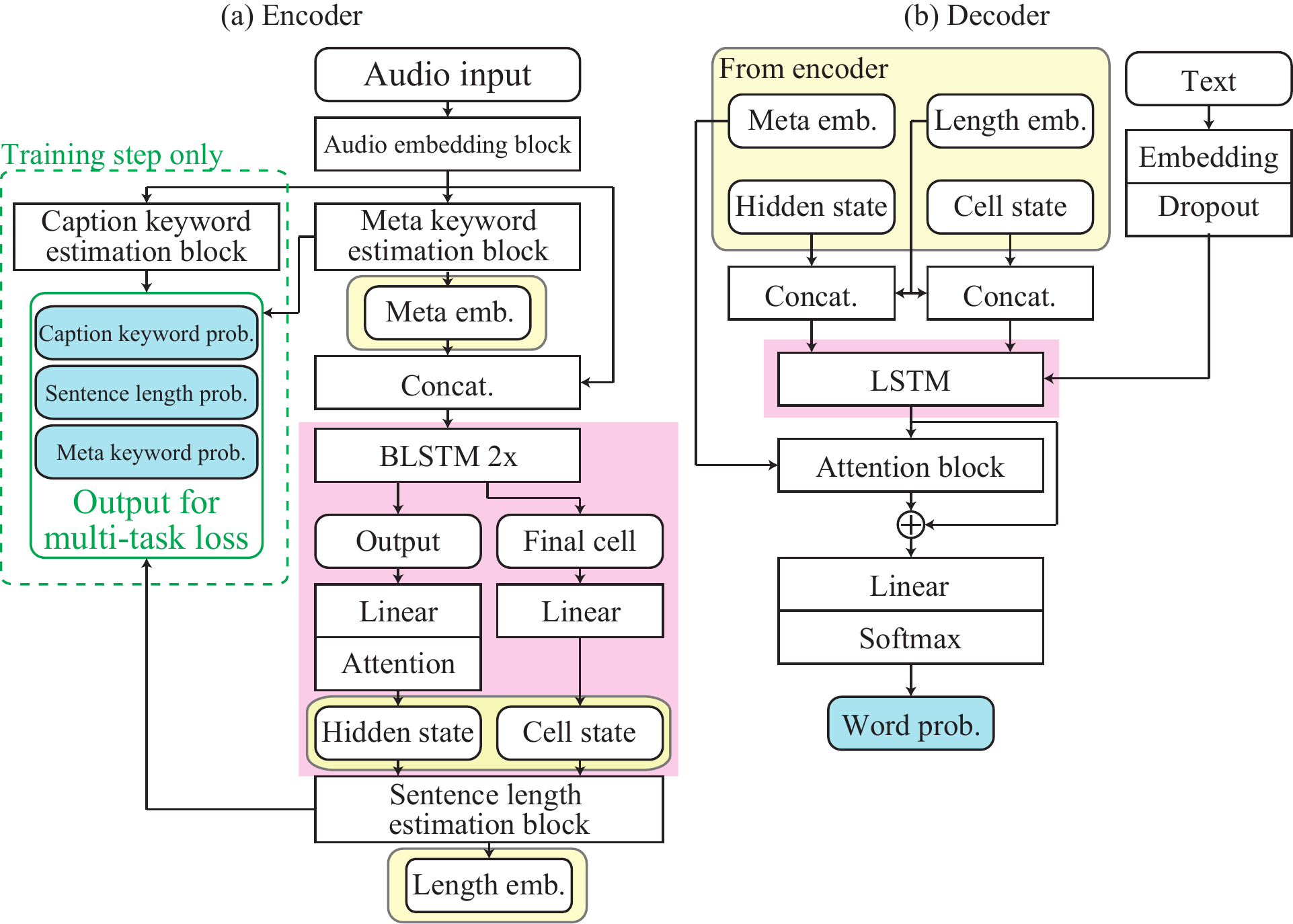} 
  \vspace{-10pt}
  \caption{Architecture of captioning DNN.}
  \label{fig:modelov}
  \vspace{-10pt}
\end{figure}

\vspace{3pt}
\noindent
{\bf Mix-up:} 
For audio augmentation, we also investigate whether the mix-up that has been validated in past DCASE challenge tasks is effective. 
Since text is input as a set of class labels, direct mixing of $\bm{w}$ is not suitable. Thus, the embedded word tokens are mixed by multiplying the mixing parameters~\cite{guo2019augmenting}.


\mysubsection{Multi-task training for resolving indeterminacy}

To solve the indeterminacy problems, we build a DNN architecture and loss functions based on multi-task learning to simultaneously estimate the caption keywords, meta keywords and sentence length.
Figure \ref{fig:modelov} shows the architecture of the captioning DNN.
The pink area in Fig.~\ref{fig:modelov} is a basic sequence-to-sequence (Seq2Seq)--based captioning model \cite{sutskever2019sequence} using bidirectional long short-term memory (BLSTM)--LSTM. 
The output and final cell state of the encoder BLSTM are used for the initial hidden state and cell state of the decoder LSTM.
Then, the decoder LSTM estimates the posterior probability of the $n$th word given the audio signal $\bm{x}$ and the 1st to $(n-1)$-th words $p( w_n | \bm{x}, w_{1,..., n-1} )$ using embedded word tokens.
In addition, we use sub-blocks for keyword and sentence length estimation to solve the indeterminacy problems in AAC. 
The following describes an overview of these sub-blocks. Please refer to our technical report~\cite{koizumi2020t6ntt} for the details of their implementations. 

\vspace{3pt}
\noindent
{\bf Meta keyword estimation:} 
The meta keyword estimation block also follows the audio embedding block, and its output is passed to the encoder. We expect this block to estimate keywords that humans associate with sounds, such as information from vision and the surrounding environment and thus avoid the indeterminacy of word selection.

In this study, the meta keywords were extracted from the \url{file_name} and \url{keyword} provided in the metadata CSV file, because they indicate their content, and a short textual description.
We extracted the keywords using a keyword vocabulary that was manually created beforehand.
The procedure for creating the keyword vocabulary is as follows. First, \url{file_name} and \url{keyword} were split at spaces and punctuation marks. Next, words that seem to be nouns, verbs, adjectives, and adverbs were converted to theirlemmas. Finally, all those were counted, and lemmas that appeared more than ten times were appended in the keyword vocabulary, which is a hash table that maps the original word to its lemma.

The meta keyword estimation block is used to avoid the indeterminacy of word selection in the same way that the caption keyword estimation block is used. 
This block estimates meta keyword probabilities from the output of the audio embedding block.
The weight of these losses are based on the prior probability of the $i$th keyword calculated by
\begin{align}
\frac{\mbox{\# of $i$th keyword in training samples}}{ \mbox{\# of training samples} }.
\label{eq:weight}
\end{align}
By using the inverse of this prior probability as the weight, the word and topic frequencies are considered in the training phase.

\vspace{3pt}
\noindent
{\bf Caption keyword estimation:}
The caption keyword estimation block follows the audio embedding block. In contrast to the meta keyword estimation block, this block is not used in the testing phase. We expect this block to guide the training of the audio embedding block so that its output includes information about the keywords of the ground-truth caption. This also results in avoiding the indeterminacy of word selection. 

This block estimates the caption keyword probabilities from the output of the audio embedding block. The caption keywords are the lemma of words frequently used in the provided meta data. The loss function for this block is the weighted binary-cross entropy. The weight of the loss is Eq.~\eqref{eq:weight}.

\vspace{3pt}
\noindent
{\bf Sentence length estimation:} 
The sentence length estimation is used to avoid the indeterminacy of sentence length. 
This block estimates sentence length probability and its embedding from the output and final cell state of the encoder BLSTM.
The sentence length embedding is passed to the decoder to generate the sentence using the length information as a part of the initial hidden state and the cell state of the decoder LSTM.

We use the softmax cross entropy as the loss for the sentence length estimation block. By using these losses for multi-task learning, we expect the common audio embedding block to be learned as a more general representation for any task and that the AAC scores are improved.

\vspace{3pt}
\noindent
{\bf Keyword co-occurrence loss:}
To prevent the decoder from outputting words obviously unrelated to the meta keywords, we additionally use the keyword co-occurrence loss between words in a caption and its meta keywords.
For example, when meta keywords are \{{\it car, sing, bird}\}, words not related to keywords such as \{{\it people, children, talking, talk, speak}\} may not be included in the correct caption.
To prevent the decoder from outputting such words, we adopt a penalty based on the decoder outputs $p( w_n | \bm{x}, w_{1,..., n-1} )$.
To implement this loss, we create a hash-table of the co-occurrence lists before training.
Then, in the training phase, we add the penalties of the decoder outputs to the whole loss value.

\begin{table*}[!t]
\vspace{-5pt}
\caption{Results of element-wise ablation study}
\label{tab:result1}
\centering
\begin{tabular}{ @{\hspace{5pt}}L{0.65}@{} | ccccccccc }
\toprule
\textbf{Model} & \textbf{B-1} & \textbf{B-2} & \textbf{B-3}	& \textbf{B-4} & \textbf{METEOR} & \textbf{ROUGE-L} & \textbf{CIDEr} & \textbf{SPICE} & \textbf{SPIDEr} \\	
\midrule
$\mathtt{Baseline \, of \, DCASE2020 \, Task6}$~\cite{dcase2020task6}
& 38.9& 13.6&  5.5& 1.5 & 26.2 & 8.4&  7.4& 3.3&  5.4 \\
\midrule 
$\mathtt{Proposed}$
& $\bm{51.2}$& 32.5& 21.5& 14.0& 34.3& 14.5& 29.0& 8.9& 19.0 \\
$\;$ w/o (i) Data augmentaion
& 52.0& 31.2& 20.0& 12.7& $\bm{33.7}$& 14.0& 26.1&$ \bm{8.2}$& $\bm{17.2}$ \\
$\;$ w/o (ii) Multi-task learning
& 51.8& 33.0& 21.7& 14.1& 34.6& 14.7& 29.1& 9.2& 19.2 \\
$\;$ w/o (iii) Post-processing
& 51.8& $\bm{30.1}$& $\bm{18.0}$& $\bm{10.8}$& 33.8& $\bm{13.9}$& $\bm{25.3}$& 9.0& $\bm{17.2}$ \\
\midrule 
$\;$ w/o all three elements 
& 52.1& 29.4& 17.4& 10.3& 33.5& 13.8& 23.2& 8.5& 15.8 \\
\bottomrule
\end{tabular}
\vspace{-5pt}

\caption{Results of module-wise ablation study}
\label{tab:result2}
\centering
\begin{tabular}{ @{\hspace{5pt}}L{0.65}@{} | ccccccccc }
\toprule
\textbf{Model} & \textbf{B-1} & \textbf{B-2} & \textbf{B-3}	& \textbf{B-4} & \textbf{METEOR} & \textbf{ROUGE-L} & \textbf{CIDEr} & \textbf{SPICE} & \textbf{SPIDEr} \\	
\midrule
(i) Data augmentation \\
$\quad$ w/o (a) Mix-up
& 51.5& 31.5& 20.4& 13.1& $\bm{33.6}$& 14.3& 27.9& $\bm{8.4}$& 18.2 \\
$\quad$ w/o (b) TF-IDF-based word placement
& 51.9& 32.6& 21.4& 13.9& 34.5& 14.6& 28.9& 8.8& 18.8 \\
$\quad$ w/o (c) IDF-based sample selection
& $\bm{51.1}$ & 32.4& 21.6& 14.2& 34.2& 14.5& 29.3& 8.9& 19.1 \\
\midrule 
(iii) Post-processing\\
$\quad$ w/o (a) Beam search decoding
& 52.1& $\bm{30.3}$& $\bm{18.1}$& $\bm{10.8}$& 33.9& $\bm{14.0}$& $\bm{25.6}$& 9.2& $\bm{17.4}$ \\
$\quad$ w/o (b)TTA
& 51.5& 32.8& 21.8& 14.3& 34.2& 14.5& 29.4& 8.9& 19.2\\
\bottomrule
\end{tabular}
\vspace{-5pt}
\end{table*}

\mysubsection{Post-processing}
We use beam search decoding for making word decisions process from $p( w_n | \bm{x}, w_{1,..., n-1} )$. The beam size is 5, and $n$-gram blocking size is 2, i.e., a hypothesis in a beam is discarded if a bi-gram appears more than once within it.
In addition, we use test time augmentation (TTA) for audio input.
This is because the audio input is randomly cropped to limit the length of time in the training phase.
If changing the length of the audio input in the testing phase could have a bad influence on the batch normalization layers.
Therefore, in the testing phase, we also randomly crop and zero-pad the audio input so that all the inputs have the same length.
We generate five input audio clips by this process, and take the average of the five outputs of the decoder.

\mysection{Experiments}
\mysubsection{Experimental setup}
As the training and test dataset, we used the Clotho dataset \cite{drossos2019clotho}, which consists of audio clips from the Freesound platform \cite{font2013freesound} and whose captions were annotated via crowdsourcing \cite{lipping2019crowdsourcing}. 
We used the development split of 2,893 audio clips with 14,465 captions (i.e., one audio clip has five ground-truth captions) for training and the evaluation split of 1,045 audio clips with 5,225 captions for testing. 
From the development split, 96 audio clips and their captions were randomly selected as the validation split. 
The setups of the DNN architecture and training were the same as in~\cite{koizumi2020t6ntt}


To evaluate our system, we use the same metrics as Task 6: BLEU-1, BLEU-2, BLEU-3, BLEU-4, ROUGE-L, METEOR, CIDEr~\cite{vedantam2015cider}, SPICE~\cite{anderson2016spice}, and SPIDEr~\cite{liu2017improved}.
Among these scores, the scores developed for captioning task are CIDEr, SPICE, and SPIDEr. CIDEr evaluates TF-IDF weighted $n$-gram similarity between the output sentence and the reference sentences, which ensures generated captions are syntactically fluent, and SPICE compares grammatically parsed generated and references sentences, which ensures generated captions are semantically faithful to the audio. SPIDEr is the linear combination of CIDEr and SPICE, and it was used as the target metric in this challenge.


\mysubsection{Experiment-I: Element-wise ablation study}
We first conducted an element-wise ablation study; the three processing elements (i) data augmentation, (ii) multi-task learning, and (iii) post-processing were excluded one each, and the accuracy with the remaining two was compared. In addition, we also evaluated the performance when all the three elements were excluded, namely, the performance of a pure seq2seq model. We trained each model three times with different initial values, and calculated the average scores. We expected this experiment to clarify which processing elements have a significant impact on accuracy and help us find the processing element to focus on in future AAC studies.


Table~\ref{tab:result1} shows the results of the element-wise ablation study. $\mathtt{Baseline}$ means the scores of the baseline model for the DCASE 2020 Challenge Task 6 \cite{drossos2017automated}, and $\mathtt{Proposed}$ is the full model used in our submission. 
These results suggest the following:

\vspace{3pt}
\noindent
{\bf Effective elements:} (i) Data augmentation and (iii) post-processing elements significantly contributed to the performance of our submission. When one of these elements was removed, the SPIDEr score decreased by 1.8 points. In addition, the SPIDEr score decreased by 3.2 points when all the elements were removed. Thus, it can be considered that the effects of these elements are independent of each other and that the combination of these elements is effective.

\vspace{3pt}
\noindent
{\bf Non-effective element:} (ii) Multi-task learning element was not very effective under this primitive rule. Surprisingly, even though this element has been proven to be effective in more computationally intensive scenarios, such as ones with pre-trained models \cite{koizumi2020transformer}, it became clear that it was not effective in our submission. Since we simultaneously trained all the DNN modules from scratch using multi-task losses, they were probably not trained effectively. For these modules to be effective, pre-training might be required, especially when the number of the trainable parameters is large, as it was in our submission.

\mysubsection{Experiment-II: Module-wise ablation study}
Next, we conducted a module-wise ablation study to assess the contribution of each module.
We calculated the scores of the models that any of the modules in processing blocks (i) and (iii) were excluded one each, where (i)~data augmentation and (iii) post-processing blocks include modules \{mix-up, TF-IDF-based word placement, IDF-based sample selection\} and \{beam search decoding, TTA\}, respectively. As in Experiment-I, we trained each model three times with different initial values, and calculated the average scores. 

Table~\ref{tab:result2} shows the results of the module-wise ablation study, and which written in the same manner of the Experiment-I. These results suggest us the following three things:

(I) The mix-up was effective, as used in acoustic event detection and acoustic scene classification.

(II) The beam search decoding worked well, as used in other text generation tasks. Especially the CIDEr score, which depends on TF-IDF, increased. 
Since the AAC system without beam search decoding maximizes the posterior probability of each word, 
most of the output word sequences is high-frequency words in the dataset, not an optimal sentence sequence. 
Therefore, it can be considered that the beam search decoding has little impact on uni-gram-based metrics such as BLEU-1 and SPICE and improves the metrics that evaluate from the perspective of the whole sentence such as BLEU-4 and CIDEr.

(III) The use of the IDF-based sample selection and TTA did not improve or worsen the performance. The reason for this may be that the parameters we used in these methods  in our implementation had little effect on the statistics of the augmented data.
In TTA, the length of cropped samples was 20 sec, although that of training/testing samples was also around 20 sec. Thus, all the multiple cropped inputs have almost the same statistics as the original one, and in which would explain why TTA did not work effectively.
In addition, in the Clotho dataset, to keep reference captions consistent, the words that appear in captions are intentionally designed lowly diverse. Therefore, simple sample or word replacement would not have been sufficient to augment training samples.
To drastically vary the statistics of the augmented data, we may need to use para-phrase generation~\cite{sokolov2020neural,lewis2020pretraining} or thesaurus-based word placement~\cite{wei2019eda}.

\label{sec:exp}

\section{Conclusions}

In this paper, we presented an audio captioning system with data augmentation, multi-task learning, and post processing, and conducted a detailed ablation study to clarify which elements are effective.
From the results, mix-up data augmentation and beam search decoding were effective in improving the accuracy of AAC.
In particular, since beam search decoding selects output from the perspective of the whole sentence,
BLEU-4 and CIDEr, which reflect n-gram accuracy, were significantly improved.
Future works include the use of data augmentation that maintains grammatical structure such as para-phrase generation for sufficient augmentation of training samples.

\clearpage
\bibliographystyle{IEEEtran}
\bibliography{refs}

\end{sloppy}
\end{document}